\def\be{\begin{equation}}
\def\ee{\end{equation}}
\def\ber{\begin{eqnarray}}
\def\eer{\end{eqnarray}}
\def\bers{\begin{eqnarray*}}
\def\eers{\end{eqnarray*}}

\def\JPC{J. Phys. C}
\def\JPCM{J. Phys.: Condens. Matter}
\def\PR{{ Phys. Rev.}\ }
\def\PRL{{ Phys. Rev. Lett.}\ }
\def\JPC{{ J. Phys. C: Solid State Phys.}\ }
\def\JPCM{{ J. Phys.: Condens. Matter}\  }

\documentclass[aps,prb,twocolumn,groupedaddress,showpacs,amsmath,amssymb]{revtex4-1}
\usepackage{dcolumn}   
\usepackage{amsmath,array}
\usepackage{graphicx}    
\usepackage{subfigure}
\usepackage{color}
\newcommand{\comment}[1]{}
\usepackage{ifthen}
\newboolean{includefigs}
\setboolean{includefigs}{true}      
\newboolean{includetext}
\setboolean{includetext}{true}     
\newcommand{\condcomment}[2]{\ifthenelse{#1}{#2}{}}
\begin{document}

\title{Accurate and fast numerical solution of Poisson's equation for arbitrary, space-filling Voronoi polyhedra: near-field corrections revisited}

\author{Aftab~Alam$^{1}$, Brian~G.~Wilson$^{2}$, and D.~D.~Johnson$^{1,3}$}
\email[emails: ]{{ddj,aftab}@ameslab.gov,wilson9@llnl.gov}
\affiliation{$^{1}$Division of Materials Science and Engineering, Ames Laboratory, Ames, Iowa 50011, USA}
\affiliation{$^2$Lawrence Livermore National Laboratory, P.O. Box 808, Livermore, California 94550, USA }
\affiliation{$^{3}$Department of Materials Science \& Engineering, Iowa State University, Ames, Iowa 50011, USA}

\begin{abstract} 
We present an accurate and rapid solution of Poisson's equation for space-filling, arbitrarily-shaped, convex Voronoi polyhedra (VP); the method is  O(N$_{{\text{VP}}}$), where N$_{{\text{VP}}}$ is the number of distinct VP representing the system.
In effect, we resolve the longstanding problem of fast but accurate numerical solution of the \emph{near-field corrections} (NFC), contributions to each VP potential from nearby VP -- typically involving multipole-type conditionally-convergent sums, or fast Fourier transforms.
Our method avoids all ill-convergent sums, is simple, accurate, efficient, and works generally, i.e., for periodic solids, molecules, or systems with disorder or imperfections. 
We demonstrate the method's practicality by numerical calculations compared to exactly solvable models.
\end{abstract} 
\date{\today}
\maketitle

\section{Introduction}
\label{sec_Intr}
{\par} Poisson's equation describes the electrostatics by relating a charge distribution to the potential contingent upon the boundary conditions. 
An accurate solution of Poisson's equation is critical in various areas of chemistry and condensed-matter physics. 
In  {\it ab initio} electronic-structure methods, the Poisson equation is solved repeatedly, and concomitantly parallel to the Schr\"odinger's equation. 
As such, computational time for solving Poisson equation is always a concern. 
Although a number of proposals exist, most suffer from  shortcomings that affect accuracy and speed, and the ability to scale to large system sizes efficiently.
Here we provide an exact treatment of Poisson's equation and its accurate and efficient numerical solution of the potential and Coulomb energy of systems described by arbitrarily-shaped, convex, space-filling VP in any site-centered method.
Our new approach scales linearly with the number of VP, and avoids mathematical and numerical issues associated with previous methods, particularly multipole approaches.
In historical context, we provide an efficient and accurate means to compute the so-called ``near-field corrections'' (NFC), a problem not fully resolved so far.

\begin{figure}[t]
\centering
\includegraphics[width=6.6cm]{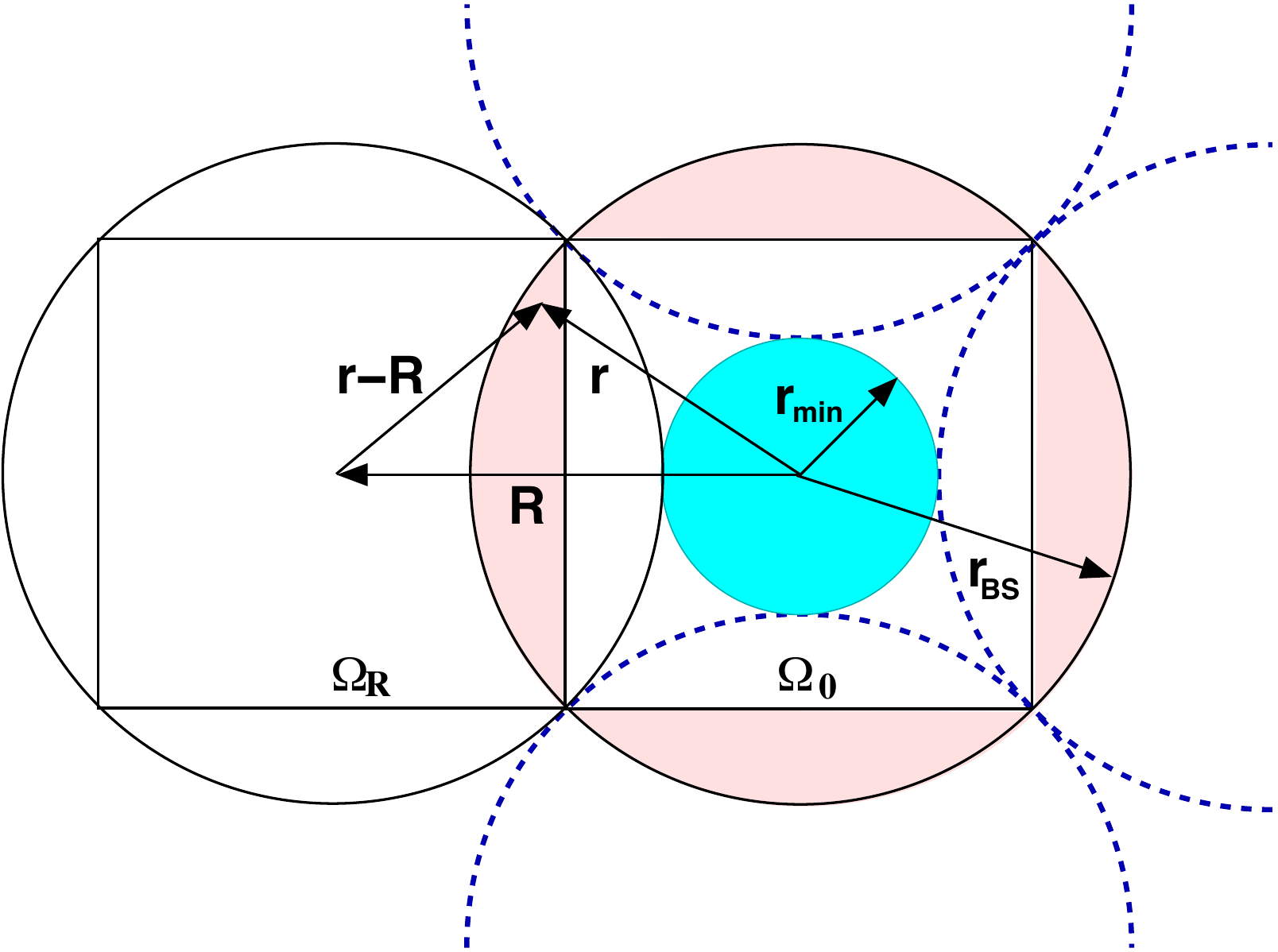}
\caption {(Color online) Two VPs ($\Omega_0$ and $\Omega_R$) separated by vector ${\bf R}$ with overlapping bounding spheres with radius ${\bf r_{BS}}$. For $r_{\text{min}} < r < r_{BS}$, NFC are needed. $r_{MT}$ is the inscribed sphere radius (not drawn for clarity).}
\label{cartoon}
\end{figure}

{\par}Typically, the electrostatic potential at a point in a convex VP is given by two contributions,\cite{Morgan77,Gonis91,Schadler92,Zhang94,Stefanou_Wang,Weinert,Vitos95,Nicholson2002,Weinberger_book} namely, (i) an intracell potential arising from the charge density within a VP ($\bar\rho^{(0)}$ in $\Omega_0$) and (ii) an intercell potential arising from all other $\bar\rho^{(R)}$ in $\Omega_R$'s, see Fig. \ref{cartoon}. 
In general,
\begin{eqnarray}
V({\bf r}) &=& \sum_{R}\int \frac{\bar\rho^{(R)}({\bf r'})~d{\bf r'} } {\vert\bf {r-(r'+R)}\vert} = V^{\text{\tiny{Intra}}}({\bf r}) + V^{\text{\tiny{Inter}}}({\bf r}), \nonumber\\ 
              &=& \int_{\Omega_{0}}  \frac{\bar\rho^{(0)}({\bf r'})~d{\bf r'}} {\vert\bf {r-r'}\vert} + \sum_{R\ne 0}\int_{\Omega_{R}}  \frac{\bar\rho^{(R)}({\bf r'})~d{\bf r'}}{\vert\bf {r-(r'+R)}\vert},
\label{eq1}
\end{eqnarray}
where $\bar\rho^{(R)}$ is a truncated density centered at site $R$.
Computational time in most methods\cite{Morgan77,Gonis91,Schadler92,Zhang94,Stefanou_Wang,Weinert,Vitos95, Nicholson2002,Weinberger_book} arise from the use of $L$~$\equiv\{l,m\}$ multipole (spherical-harmonics $Y_{L}(\widehat{\bf{r}})$) expansions.
Evaluation of intercell potential (term two in Eq.~\eqref{eq1}) is the most tricky, and our main focus. 
Often, as a first step, the Green's function $\vert {\bf {r-(r'+R)}}\vert^{-1}$ is expanded in $Y_{L}$'s in terms of $r_{<}$ (e.g., $\vert {\bf r}\vert$) and $r_{>}$ (e.g., $\vert {\bf {r'+R}}\vert$), see Sec. \ref{sec_method}, attempting to separate two of three ($r, r', R$) degrees of freedom. 
In most existing methods,\cite{Morgan77,Gonis91,Schadler92,Zhang94,Stefanou_Wang,Weinert,Vitos95, Nicholson2002,Weinberger_book} an additional multipole expansion of $Y_{L}(\widehat{\bf{r'+R}})$ is performed yielding {\it conditionally-convergent} nested $L$-sums (internal vs. external: $l^{\text{int}}_{\text{max}} > 3l^{\text{ext}}_{\text{max}}$) due to the nearest-neighbor sites, and relevant in the light shaded (pink) region in Fig.~\ref{cartoon}. 
Such nested sums are numerically expensive and ill convergent, even more so for distorted (asymmetric) cells. 
Numerical inefficiency also arises from any use of VP {\it shape functions},\cite{Morgan77,Stefanou_Wang} which utilize $Y_{L}$'s to expand VP shapes to facilitate VP integrations; again, these are costly (and inaccurate) due to the large $L$-sums ($l^{\text{int}}$$>>$$3l^{\text{ext}}_{\text{max}}$) required.
For ``muffin-tin'' potentials varying only inside $r_{MT}$ (Fig.~\ref{cartoon}), these issues are moot as no conditional expansions are needed; the ``atomic sphere approximation'' ignores these errors.

{\par}Thus, for arbitrarily-shaped, convex, space-filling VP, we derive the set of integral equations that permit us to eliminate all previous computational bottlenecks and convergence issues to solve Poisson's equation by employing isoparametric integration,\cite{Alam2011} valid for rapidly varying and/or decaying integrands, while providing a dramatic savings of computational time, \emph{e.g.}, $10^5$ in time and $10^7$ in accuracy over the shape-functions!
The method permits site-specific quantities to be calculated rapidly, scales linearly with the number of VP N$_{\text{VP}}$ and is easily parallelized.
Unlike the Full-potential Linear Augmented Plane-Wave (FLAPW) method, Fast Fourier Transforms (FFTs), which   limit scaling to large systems, are not needed.
To prove these points explicitly, we compute example integrals for potential and Coulomb energy from analytic charge-density models.\cite{Morgan77,Slater1967} 

\section{Background} 
\label{sec_Backgrd}

{\par}To solve Poisson's equation for site-centered methods, various techniques have been developed.
Gonis {\emph {et al.}}\cite{Gonis91} introduced a technique (modified later by Vitos {\emph {et al.}}\cite{Vitos95}) based on shifting (and back-shifting) the neighboring cells by a vector {\bf b} that eliminates the conditionally-convergent expansion related to these neighbors, but requires additional $L$ sums; the technique converges very slowly versus $L_{max}$ because internal sums are large, e.g., $l^{\text{int}}_{\text{max}} > 3l^{\text{ext}}_{\text{max}}$; additionally, {\bf b} is a parameter that must be chosen wisely and depends on crystal symmetry. 
Others\cite{Stefanou_Wang} used shape-functions making the VP integrations very fast for a $Y_{L}$-basis but the expansion is slowly convergent (i.e., $l^{\text{int}}_{\text{max}}>30$), with limited accuracy.\cite{Alam2011}
Schadler \cite{Schadler92} proposed corrections to the usual multipole expansion via a conditionally-convergent formula due to Sack;\cite{Sack1964} however, these corrections do not satisfy Laplace's equation.  
Zhang {\emph {et al.}}\cite{Zhang94} converted VP integrals to surface integrals, avoiding most conditionally-convergent sums; however, it is not automated for complex geometries, and concerns remain about degeneracies for their set of linear equations.
For FLAPW, Weinert\cite{Weinert} avoided these issues via $Y_{L}$-basis in MT-spheres and interstitial plane-waves; however, to obtain a smooth density (for a chosen set of MT radii) a large number of plane waves (N$_{\text{PW}}$$>$30,000) and $Y_{L}$'s ($l_{\text{max}}\ge8$) are required, and one never obtains VP-specific properties.
FFTs are then needed, scaling as 2N$_{\text{PW}} log$(N$_{\text{PW}}$), with specialized programming for large system sizes. 
For Linear Combination of Atomic Orbital (LCAO) methods,\cite{NRLmol,ADF} various atomic bases (e.g., Gaussian orbitals) are used in different regions of space to study molecules and clusters. 
Gaussian-orbital methods do not necessarily require partitioning of space because Poisson's equation can be solved analytically (or in terms of incomplete Gamma functions) on any mesh of points. 
However, a significant advantage could be achieved by a method that solves Poisson's equation numerically and accurately; for example, some Gaussian-orbital codes resort to least-square fits to solve Poisson's equation because it is faster albeit approximate.\cite{Dunlap}

\section{A computationally efficient and accurate Poisson solver} 
\label{sec_method}

{\par} A proposal by Nicholson and Shelton\cite{Nicholson2002} is conceptually easy, although it suffers also from convergence issues -- both multipoles and shape-functions.
We use a key idea from their work but, uniquely in our derivation, avoid any expansions used in prior approaches, made possible by isoparametric integration.\cite{Alam2011}

{\par}To start, using $L~\equiv\{l,m\}$ as a composite index, we express the solution of Poisson's equation as\cite{Zhang94}
\begin{eqnarray}\label{eq2}
V({\bf r}) &=& \sum_{L}^{l_{\text{max}}}  \left[V_{L}^{\text{ex}}(r) + \alpha_{L}\ r^l\right] Y_{L}(\widehat{\bf {r}}), \ r\le r_{BS} \ \ \ \ \ \    \\
\text{ with~} V_{L}^{\text{ex}}(r)&=&\frac{4\pi}{2l+1}\left[ w_{L}(r) + r^{l} \int_{r}^{r_{BS}} dx\ \frac{\rho_{L}^{\text{ex}} (x)}{x^{l-1}}\right] . \nonumber
\end{eqnarray}
$\rho_{L}^{\text{ex}} (r)$ is the extended charge density inside the circumscribing (or bounding) sphere of radius $r_{BS}$ of the central cell $\Omega_0$ in  Fig.~\ref{cartoon}. 
The radial function $w_{L}(r)$ is the contribution to the potential within a distance $r$ from origin of $\Omega_0$, which is given by
\begin{equation}\label{eq2b}
w_{L}(r)=r^{-(l+1)} \int_{0}^{r} dx~x^{l+2} \rho_{L}^{\text{ex}}(x)    , 
\end{equation}
and which is bounded, i.e., $w_{L}(r \rightarrow 0) = 0$, and finite for any $r \le r_{BS}$, and, therefore, easily integrated.

The intracell potential is the first term in Eq.~\eqref{eq2}, while the intercell potential was expressed as $\alpha_{L} r^l Y_{L}$ to make apparent a mathematical ``trick'' (assignment of equality) used below.
Here $\alpha_{L}$ is an unknown coefficient depending on the charge distribution of the system. 
The main objective is to determine $\alpha_{L}$, which, if known, would give the potential at any point inside the central $r_{BS}$ sphere.

{\par}The problem in calculating $V^{\text{Inter}}({\bf r})$ directly in Eq.~\eqref{eq1} is the need to assume (particularly for multipole approaches) the geometric condition
\begin{eqnarray}\label{eq3}
r<\vert{\bf r'+R}\vert, \ \ r'<R,
\end{eqnarray}
which is not fulfilled in the so-called {\it moon region} between the near VP cells,\cite{Gonis91,Schadler92,Weinberger_book}  shown by light (pink) shading in Fig.~\ref{cartoon}, or, in other words, the complement of the VP and its bounding sphere with radius $r_{BS}$. 
A cell centered at ${\bf R}$ is a near-cell of the central one if $R<r_{BS}^{(0)} + r_{BS}^{(R)}$. 
Incorrect contributions to the potential arise from near VP beyond a radius $r_{\text{min}}$, which have been often ignored or badly approximated. 
If, however, we limit ourselves to  $r\le r_{\text{min}}$ (Fig. \ref{cartoon}), the geometric condition Eq.~\eqref{eq3} is valid and the potential \eqref{eq1} can be calculated easily. 
The unknown coefficients $\alpha_{L}$ can be then determined by equating Eqs.~\eqref{eq1} and \eqref{eq2} within  $r_{\text{min}}$.

{\par} Now, following this line of reasoning, with $r\le r_{\text{min}} \Rightarrow r\le \vert {\bf r'+R}\vert$, term two of Eq.~\eqref{eq1} can be expressed as\cite{Weinberger_book} 
\begin{equation}
   \sum_{L} a_L r^l Y_{L} (\widehat {\bf r}), \ \text{where} \\
\end{equation}
\begin{equation}
\label{eq4}
a_L =\sum_{R\ne 0} \frac{4\pi}{2l+1} \int_{\Omega_R} d{\bf r'} \bar\rho^{(R)}({\bf r'}) \frac{Y_{L}^{*} (\widehat {\bf{r'+R}})}{\vert {\bf{r'+R}}\vert^{l+1}}
\end{equation}
Rapidly varying and/or decaying integrand, as in Eq. \eqref{eq4}, over general VP can be calculated accurately and fast with an isoparametric numerical quadrature method\cite{Alam2011} with analytically-known points and weight.
(Other methods\cite{Zhang94,Weinberger_book} for performing integrals also works well, albeit not as efficiently).
A critical side point: no expansion (or FFT) of the integrand in Eq.~\eqref{eq4} is necessary, eliminating all previous computational bottlenecks and convergence issues.
A rigorous example is provided in Sec.~\ref{sec_results}.

{\par}Then, with $\bar{\rho}\rightarrow \rho^{\text{ex}}$ for $r\le r_{\text{min}}$ (the spherically symmetric regime), the first term of Eq.~\eqref{eq1} is simplified as
\begin{equation}\label{eq1.t1}
\sum_{L}\frac{4\pi}{2l+1}\left[ w_{L}(r) + r^{l} \int_{r}^{r_{BS}}d{\bf x}~\frac{\bar{\rho}^{(0)} ({\bf x})}{x^{l+1}}\ \right] 
\end{equation}
Substituting Eqs.~\eqref{eq1.t1} and ~\eqref{eq2b} into Eq.~\eqref{eq1} and comparing it with Eq.~\eqref{eq2} yields $\alpha_{L}$ for all $r_{\text{min}} \le r' \le r_{BS}$ (the remaining space), i.e.,
\begin{equation}\label{eq5}
\alpha_L = a_L + \frac{4\pi}{2l+1} \int_{\Omega_0} d{\bf x}\left[ \bar{\rho}^{(0)} ({\bf x}) - \rho^{\text{ex}} ({\bf x}) \right] \frac{ Y_L^{*} (\widehat{\bf x})}{x^{l+1}} . 
\end{equation}
\emph{Equation~\eqref{eq5} is our central result.} 
It serves to calculate accurately V$^{\text{Inter}}(\bf{r})$ with the necessary NFC, given by the integral term. 
This NFC is non-zero only beyond $r_{\text{min}}$ ($\bar{\rho}\rightarrow \rho^{\text{ex}}$ for $r\le r_{\text{min}}$) and pronounced in the ``moon region'' (${\bf r}\not\in \Omega_0~\text{and}~|{\bf r}| \le r_{BS}~\text{of}~\Omega_0$).

{\par}Notably, knowing V$^{\text{Inter}}(r<r_{\text{min}})$ gives $\alpha_L$ and, thus, V$(\bf{r})$ everywhere in space via Eq.~\eqref{eq2}, which is ultimately the ``trick''.
Finally, the cell integrations in Eq.~\eqref{eq5}, which can exhibit rapidly varying and/or decaying integrands, needs to be performed by an accurate and fast integration method over arbitrarily-shaped VP, which is satisfied by a recently proposed isoparametric integration.\cite{Alam2011}

{\par}NFC provide the correct $V^{\text{Inter}}({\bf r})$ from the near-cells, and are the motivation behind previous methods.\cite{Gonis91,Schadler92,Zhang94,Nicholson2002,Vitos95,Weinberger_book} 
Unlike existing schemes that address NFC, our derivation is simple and provides an efficient, fast and accurate solution of Poisson's equation.

{\par} In historical context, the ill-convergent sums in other methods arise from traditionally expanding $Y_{L}^{*} (\widehat {\bf{r'+R}})/\vert {\bf{r'+R}}\vert^{l+1}$ in Eq.~\eqref{eq4}, i.e., for all $r'<R$,
\begin{widetext}
\begin{eqnarray}\label{eq5a}
\frac{Y_{L}(\widehat {\bf{r'+R}})}{\vert {\bf{r'+R}}\vert^{l+1}}
= \frac{(r')^l}{R^{l+1}}\  \sum_{L^{\text{int}}} \frac{(-1)^{l^{\text{int}}-1}}{R^{l^{\text{int}}}} \frac{4\pi[2(l+l^{\text{int}})-1]!!}{(2l-1)!! (2l^{\text{int}}+1)!!}\ C_{lm,(l+l^{\text{int}})(m^{\text{int}}-m)}^{l^{\text{int}}m^{\text{int}}} \ Y_{l^{\text{int}}m^{\text{int}}}(\widehat {\bf{r'}})\ Y_{(l+l^{\text{int}})(m^{\text{int}}-m)}(\widehat {\bf{R}})
\end{eqnarray}
\end{widetext}
which separate ${\bf r'}$ and ${\bf R}$ creating a multipole-type expression via Eq.~\eqref{eq4} with large internal, conditionally-convergent sums ($L^{\text{int}}$).
The convergence of such expansions (involving Gaunt coefficients $C_{LL^{''}}^{L'}$) is sensitive to the location of ${\bf r'}$ when ${\bf R}$ is a near-cell vector, being especially difficult to converge if ${\bf r'}$ lies, e.g., near one of the corners of the VP. 
To  achieve a minimal level of convergence (e.g., $10^{-4}$), the number of $L$'s required is huge ($l>70$) even for highly symmetric VP, such as fcc and bcc!  
These errors are often ignored.

{\par} For completeness, we note that the expansion necessary for the electrostatic potential for general charge distributions in terms of spherical harmonics, like Eq.~\eqref{eq5a} has a long history which continues. 
For example, for one- and two-center Coulomb potentials, Buehler addressed spherical distributions,\cite{Buehler1951} and Fontana addressed discrete distributions,\cite{Fontana1961} Jansen provided a tensor formalism for multipole expansions;\cite{Jansen1958} however, Sack's results are well-known, as discussed in the Background section,\cite{Sack1964} and often revisited\cite{Dixon1973,vanRij1973} because of the use of hypergeometric functions, which even Sack did later.\cite{Sack1974}
Nonetheless, all the results have extensive sums that are conditionally convergent.

{\par}Finally, Gonis {\emph {et al.}}\cite{Gonis91,Vitos95} acknowledged that, in their method for solving Poisson equation, the $l$-convergence depends sensitively on the choice of the shifting vector $\bf b$ that mathematically moves the central site $\Omega_0$ far enough away from the remaining nearest-neighbor sites such that the usual $r_{<}$ and $r_{>}$ spherical harmonic expansions are valid for all ${\bf r}$ within $\Omega_0$; however, such a shifted expansion requires a very large internal $L$ sum for full convergence. 
In the resulting equations\cite{Gonis91,Vitos95,Sack1974} the shifting vector adds another conditionally-convergent summation, with multiply nested $L$ sums.
For large $l$'s, convergence further suffers due to the non-vanishing high $l_{\text{in}}$ multipole moments constructed from the shape function, giving slowly convergent inner sums for near cells and high $l_{\text{out}}$. 
Our method is free from such issues.

\section{Results and Discussion}
\label{sec_results}
{\par} To illustrate the accuracy of our method, we present results for two distinctly different cases.
First, an electronic charge density model by van W. Morgan,\cite{Morgan77} in which all results can be derived and evaluated analytically, and which mirrors the collective densities of real atoms.
Second, we address the well-known``Madelung'' problem (a jellium-like model), which has a closed-form solution using Ewald's method, but requires numerical evaluation due to appearance of non-elementary special functions (error functions), as detailed over decades and presented in Slater's book\cite{Slater1967} from the work of Slater and de Cicco.\cite{Slater1963}

\begin{table}[b]
\caption{\label{table1} $\alpha_{lm}$ calculated via Eq.~\eqref{eq5} for fcc ($R$ is summed to $8^{th}$ neighbor shell). $\{{\text{N}}_{\text{G}}\}$ is the number of Gauss points per $x,y,z$ direction for $6$ decimal place accuracy. $\alpha_{00}$ does not match the exact result due to an overall constant of integration, which depends on the crystal symmetry under consideration; however, it does not affect r-dependence.}
\begin{ruledtabular}
\begin{tabular}{lllll}
\vspace{0.2cm}
$l$  &  $m$  &  $\{{\text{N}}_{\text{G}}\}$ & $~~~~ [\alpha_{lm}]_{\text{numerical}}$ &  $~~~~ [\alpha_{lm}]_{\text{exact}}$ \\ 
\hline
\vspace{0.15cm}
0  &  0  &   12 &  ~~~2.819719207    &  ~~~2.004395351    \\
4  &  0  &   14 & ~~-6.750329999     &  ~~-6.750337649     \\
\vspace{0.15cm}     
4  &  4  &   14 & ~~-4.034089224     &  ~~-4.034098340     \\
6  &  0  &   16 & ~~-8.529479219      & ~~-8.529486709     \\
\vspace{0.15cm}
6  &  4  &   16 &  ~~15.957205113    &  ~~15.957208482    \\
8  &  0  &   18 &  ~~~4.330472442    &  ~~~4.330470922    \\
8  &  4  &   18 &  ~~~1.628477265    &  ~~~1.628476693    \\
\vspace{0.15cm}
8  &  8  &   19 &  ~~~2.481186231    &  ~~~2.481185360    \\
10 &  0  &   21 &  ~~~3.017387898    &  ~~~3.017379144    \\
10 &  4  &   21 & ~~-3.040510248     & ~~-3.040501162     \\
10 &  8  &   24 & ~~-3.618928431     & ~~-3.618920239     
\end{tabular}
\end{ruledtabular}
\end{table}

\subsection{van Morgan density model}
\label{sec_results_vanMorgan}
{\par} To illustrate the accuracy of our method for the potential and Coulomb energy, we chose an analytic model by van W. Morgan,\cite{Morgan77} whose charge density is given by
\begin{equation}
\rho({\bf r}) = B \sum_{n} e^{i {\bf T}_n . {\bf r}} \ \ .
\label{eq6}
\end{equation}
$B$ is an arbitrary constant (set to $1$) and ${\bf T}_n$ (with magnitude $|\text{T}|$) are reciprocal-lattice vectors of the system under consideration, see Ref.~\onlinecite{Alam2011} for more details with the  derived expression given in its appendix.
The exact potential for such a charge distribution is 
\begin{equation}\label{eq-vMpot}
  V({\bf r}) = 4\pi \rho({\bf r})|T|^{-2} + V_0  ,
\end{equation}
where $V_0$ is an arbitrary constant.
Also, the Coulomb energy for VP unit-cell volume $\Omega_0$ is 
\begin{eqnarray}   \label{eq7}
U &=& \frac{1}{2}\int_{\Omega_0}  \rho({\bf r}) V({\bf r}) d{\bf r} 
       \xrightarrow[]{\text{exact}} \frac{2\pi {\Omega_0}}{{|T|^{2}}}\sum_{n} 1 \ \ . 
\end{eqnarray}
This charge-density model, which mimics real (collective atomic-centered density) behavior provides a rigorous (exact) test, not possible in applications to a ``real'' system.

\begin{figure}[h!]
\centering
\includegraphics[width=6cm]{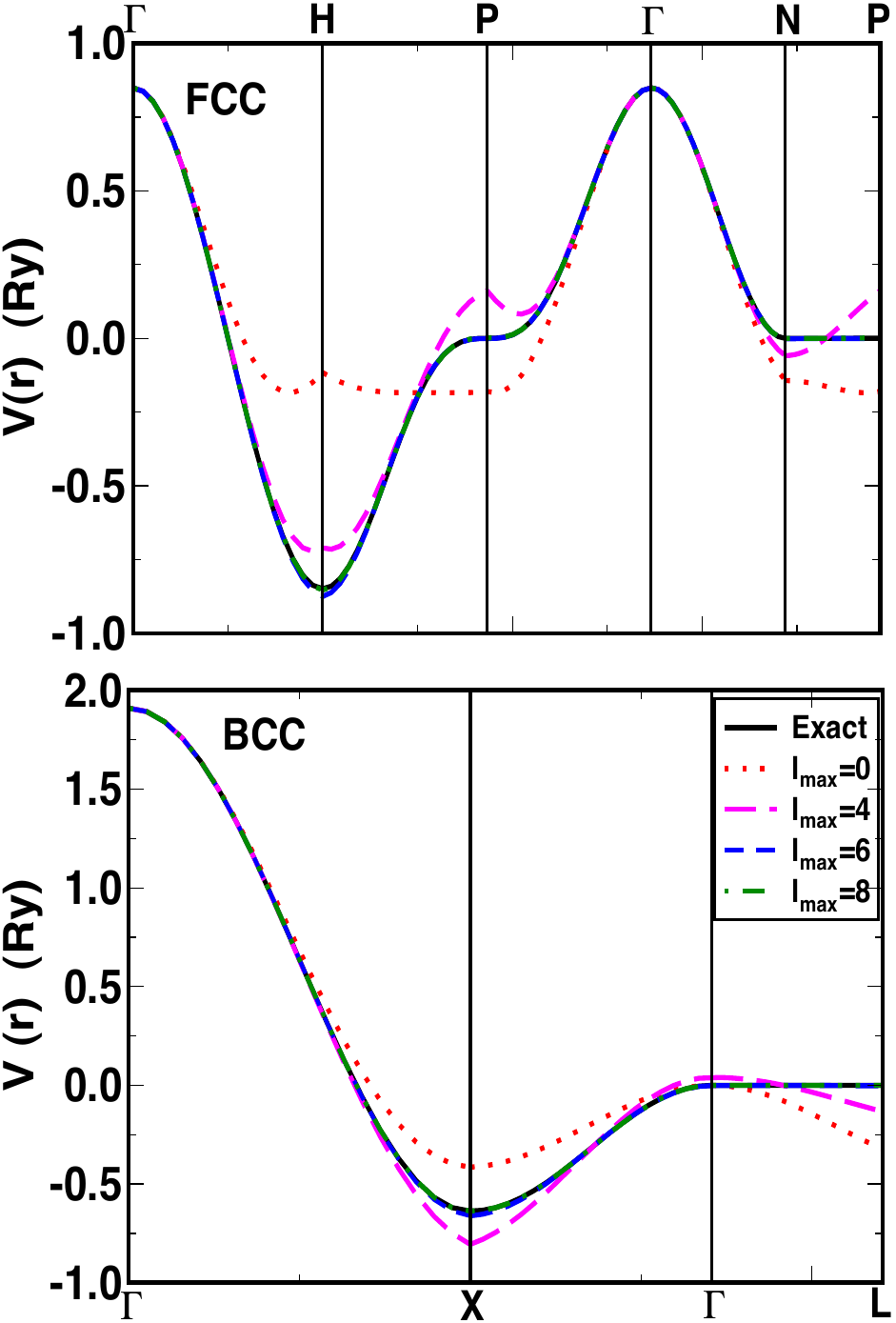}
\caption {(Color online) V({\bf r}), relative to a constant, for various $l_{\text{max}}$ along high-symmetry directions in WS-cells of fcc (top) and bcc (bottom) for van W. Morgan model.}
\label{Vr_vs_l}
\end{figure}

For the density given by Eq.~\eqref{eq6}, we evaluate the first key integral quantity, provided in Eq.~\eqref{eq5}. 
Table~\ref{table1} shows the coefficients $\alpha_L$ (Eq.~\eqref{eq5}) with respect to the number of Gauss points \{N$_{\text{G}}$\} to achieve $6$ decimal place accuracy for various $L\equiv\{l,m\}$. 
The numerically calculated $\alpha_L$ are compared with the analytical exact expression (right most column in Table~\ref{table1}) given by, with $C_{lm} = 4\pi~i^{l} \sum_{n} Y_{lm}(\widehat{T_n})$,
\begin{eqnarray}   
\alpha_{lm} &=& \frac{4\pi~j_{l-1}(|T|r_{{BS}})}{(2l+1)|T|r_{BS}^{l-1}} C_{lm}+\sqrt{4 \pi}\ V_0 \delta_{l0}
\end{eqnarray}
and $j_{l}$ are the spherical Bessel function.    
In spite of the oscillatory angular dependence in Eq.~\eqref{eq4}, with $l$-dependent spatial decay, the increase in ${\text{N}}_{\text{G}}$ required with larger $l$'s is not significant, and, hence, the isoparametric integration method used remains fast.
Only the $\alpha_{00}$ coefficient is not produced correctly, see Table~\ref{table1}; however, we note that (1) $\alpha_{00}$ is highly sensitive to the boundary conditions in the $r\rightarrow\infty$ limit and how this limit is taken, see discussion by van W. Morgan (appendix),\cite{Morgan77} or by Leeuw,\cite{Leeuw1980} which nonetheless can be solved by standard Ewald techniques; and (2) the potential is defined up to an arbitrary constant generally, as used in most electronic-structure codes to advantage. 
Hence, the error in $\alpha_{00}$ does not impact the key spatial-dependence of the potential required.

\begin{figure}[]
\centering
\includegraphics[width=5cm]{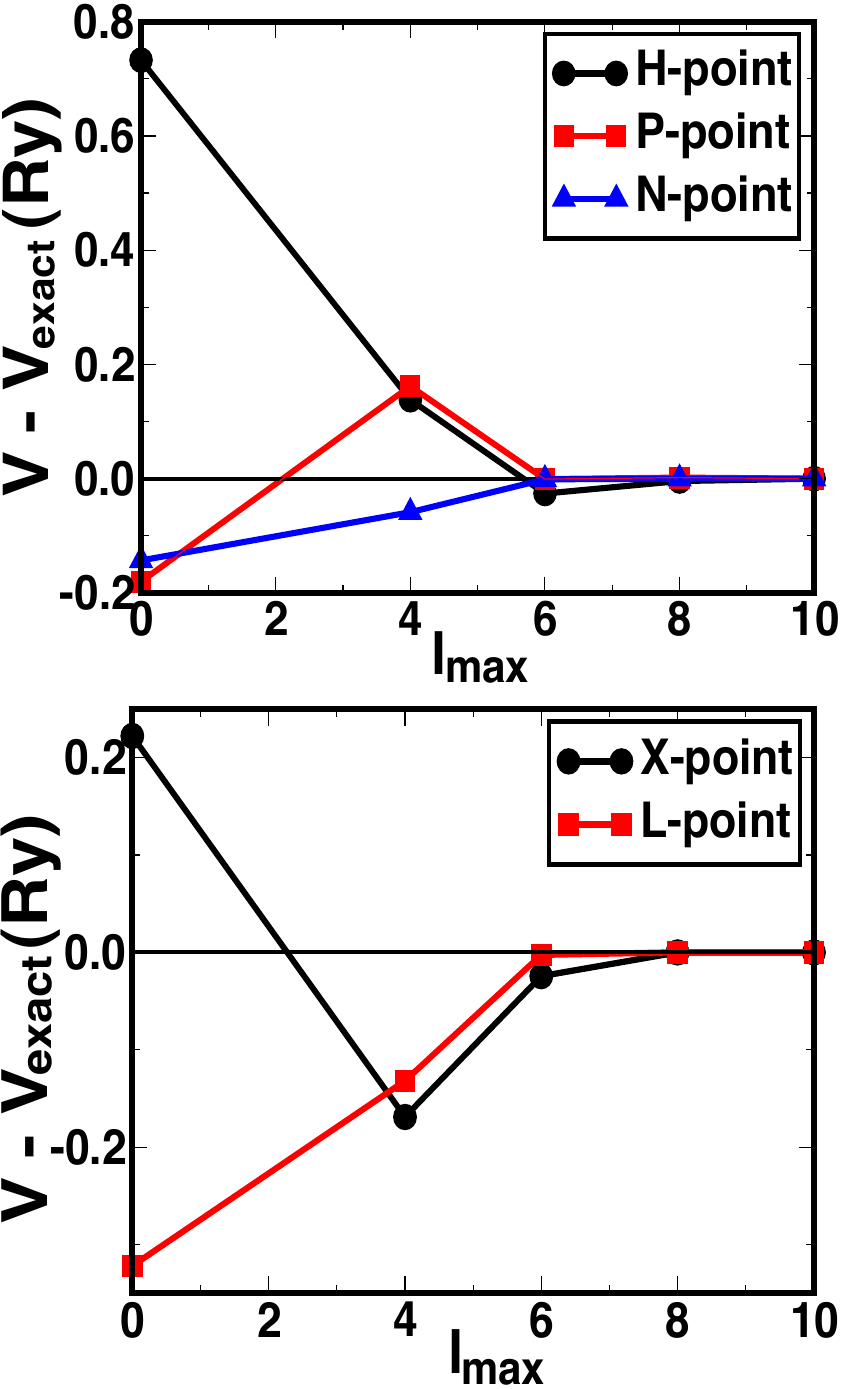}
\caption {(Color online) V({\bf r})  vs. $l_{\text{max}}$ at high-symmetry points in fcc (top) and bcc (bottom) cells for van W. Morgan model.}
\label{Vr_vs_lmax}
\end{figure}

{\par} In Fig. \ref{Vr_vs_l}, we compare $V({\bf r})$ calculated from Eq.~\eqref{eq2} for $l_{\text{max}}=0,4,6,8,10$ with that of the exact result for fcc and bcc lattices. 
The potential converges rapidly in $l$, with $l=8$ results agreeing well with  $V_{\text{exact}}$. 
The quality of agreement between the curves depends on the direction inside the VP cell, with $l$-convergence slower for points near cell boundaries. 
For instance, H (P) symmetry point is the near (far) part of the fcc VP, and X (L) is near (far) part of the bcc VP. 
Figure \ref{Vr_vs_lmax} shows the convergence of the potential at these symmetry points versus $l_{\text{max}}$; the potential at $l_{\text{max}}=6$ already converges within 0.1\% of the exact result. 
Unlike previous approaches, our method requires just one converged $L$-sum ($l_{\text{max}}\simeq 6-8$), giving a significant speed up.

\begin{figure}[b]
\centering
\includegraphics[width=7.5cm]{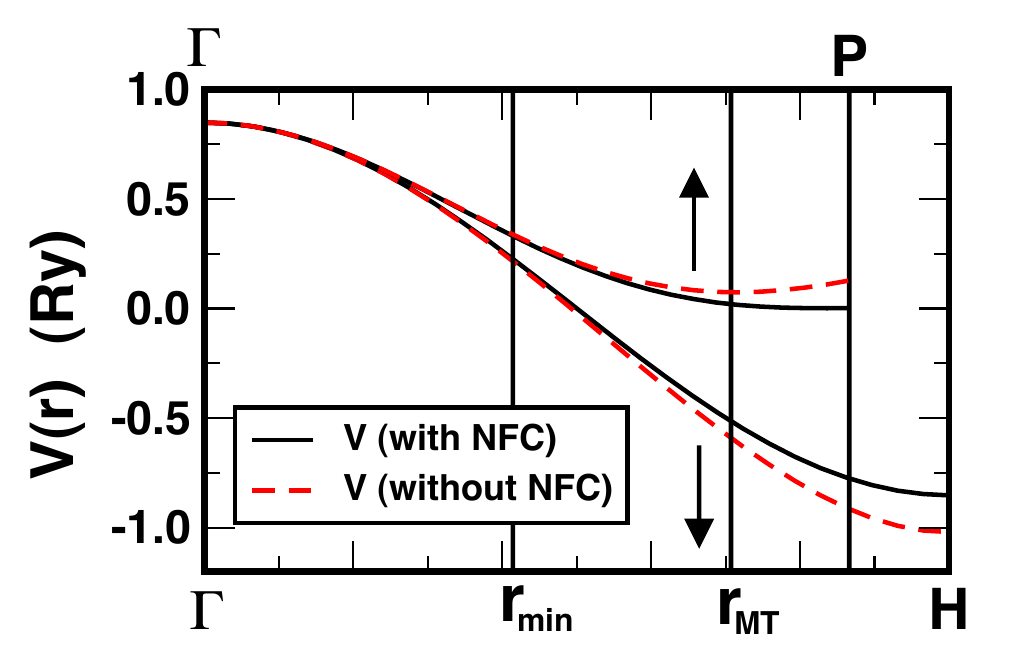}
\caption {(Color online) For fcc, the potential with (without) NFC along $\Gamma-$H and $\Gamma-$P for van W. Morgan model. Solid curves match with the exact results. $r_{\text{MT}}$ is an inscribed MT-sphere radius.}
\label{NFC}
\end{figure}

{\par} The slower rate of $l$-convergence near the cell boundary mainly arise  due to larger NFC (integral term in Eq. \eqref{eq5}) in this region, see Fig.~\ref{NFC}, where the NFC to the potential for an fcc lattice are shown along the two symmetry directions with $l_{\text{max}}=8$. 
The potential within $r_{\text{min}}$ with(out) NFC are the same as the exact result, as expected, and only beyond $r_{\text{min}}$  does the correction grow. 
The NFC, although apparently small, are very important in getting the correct result, and are larger in less-symmetric structures, which may require a higher $L$-sum to converge.
Moreover, the NFC for high $L$'s are actually very large but compensated by the $a_L$ coefficients, and, at small $L$'s the NFC are similar in magnitude to the $a_L$'s in most cases, making the integral term in Eq.~\eqref{eq5} critical to achieve the correct result.

\begin{figure}[t]
\centering
\includegraphics[width=7.5cm]{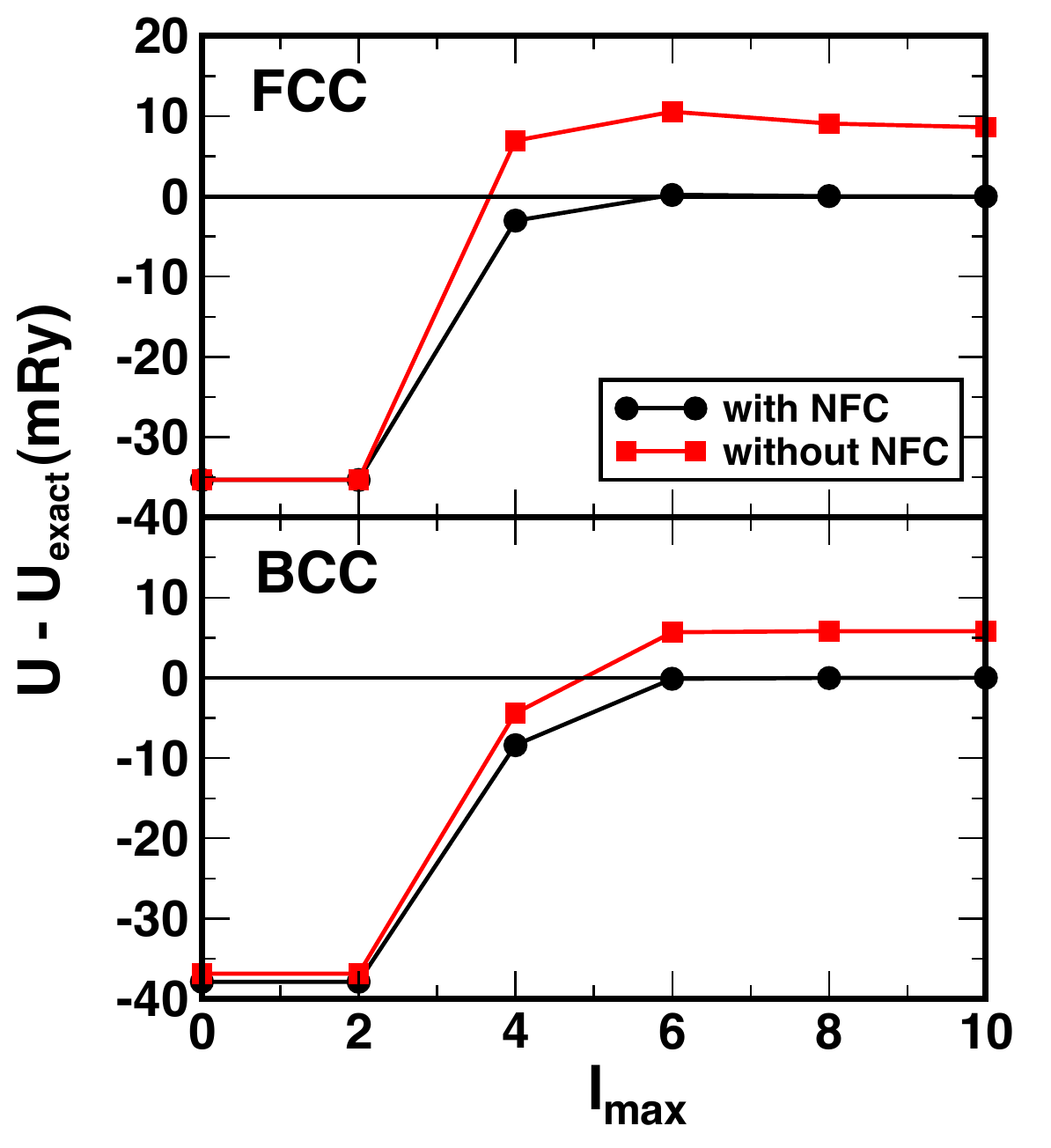}
\caption{(Color online) Coulomb energy versus $l_{\text{max}}$ for fcc (top) and bcc (bottom) lattice, with(out) NFC.}
\label{Coulomb_vs_l}
\end{figure}

{\par}Figure \ref{Coulomb_vs_l} shows the convergence of Coulomb energy $U$ versus $l_{\text{max}}$ for fcc and bcc lattices, compared to the van W. Morgan exact result. 
Without the NFC, the error is $\simeq$$10~m$Ry for fcc and $\simeq$$6~m$Ry for bcc cases, and do not improve with higher $L$'s. 
(No systematic error cancellation is possible, e.g., for U$_{\text{fcc}}$-U$_{\text{bcc}}$.)  
Unlike the potential, the Coulomb energy is almost exact by $l_{\text{max}}=6$, because $V-V_{\text{exact}}$ oscillates about zero for a given $\bf{r}$ as a function of ($\theta,\phi$) and these contributions mostly cancel when integrated over the VP, which may be true for most cases.

\subsection{Madelung's Problem}
\label{sec_results_jellium}
{\par} The Madelung ``jellium'' model consists of a constant electronic (negative) charge density throughout space, $-\rho_{0}$ ($\rho_0 = Z/\Omega_0$) which integrates to $-Z$, compensated by an ordered array of positive nuclear point charges $+Z$ at atom-center positions ${\bf R}_n$, providing charge neutrality on average, locally (within a Voronoi or Wigner-Seitz cell) and globally.
The total density then is
\begin{equation}   \label{eq-jellium-density}
\rho_{tot} = Z \sum_n \delta({\bf r} - {\bf R}_n) - \rho_0 .
\end{equation}
Via the Ewald method\cite{Ewald1921} a compensating set of positive and negative Gaussian charge distributions are used, i.e.,
\begin{equation}   \label{eq-Ewald-density}
\rho^G_i({\bf r}) = \frac{Z \epsilon^3}{\pi^{3/2}} e^{-\epsilon^2 r^2} .
\end{equation}
This extra distribution acts like an ionic atmosphere to screen the interactions between neighboring charges, which make these interactions now short-ranged, but all the Gaussian images must be summed to infinity.
A closed-form solution\cite{Slater1967} for the potential is given by
\begin{eqnarray}   \label{eq-jellium-pot}
V({\bf r}) =&& 2Z \left( \frac{4\pi}{\Omega_0} \sum_{ {{\bf K}_{m} \ne 0}} 
                         \frac{e^{-\frac{|{\bf K}_{m}|^2}{4\epsilon^2}} 
                         e^{i {\bf K}_{m}\cdot {\bf r}} }{|{\bf K}_{m}|^2}   \right)  \\
            &+& 2Z \left(  \sum_{R} \frac{ \emph{erfc}(\epsilon|{\bf r-R}|)}{ |{\bf r-R}|}
                   \right)  
            - \frac{ {2Z} \pi}{\Omega_0 \epsilon^2 }~+~V_0,   \nonumber
\end{eqnarray}
where $V_0$ is an arbitrary constant and $\epsilon$ is the Ewald parameter (controlling the width of the Gaussian in Eq.~\eqref{eq-jellium-density}), famously used to optimize the convergence of the sum used for screening, where part is done in real-space and part in k-space. 
Besides the on-site Gaussian, the \emph{erfc} function requires summation over Gaussian tails contributing from neighboring sites, however many are non-zero.
It can be verified that, with the constant of integration above, the potential is independent of $\epsilon$, as required, i.e., the first derivative with respect to $\epsilon$ is zero.

\begin{figure}[t]
\centering
\includegraphics[width=5.2cm]{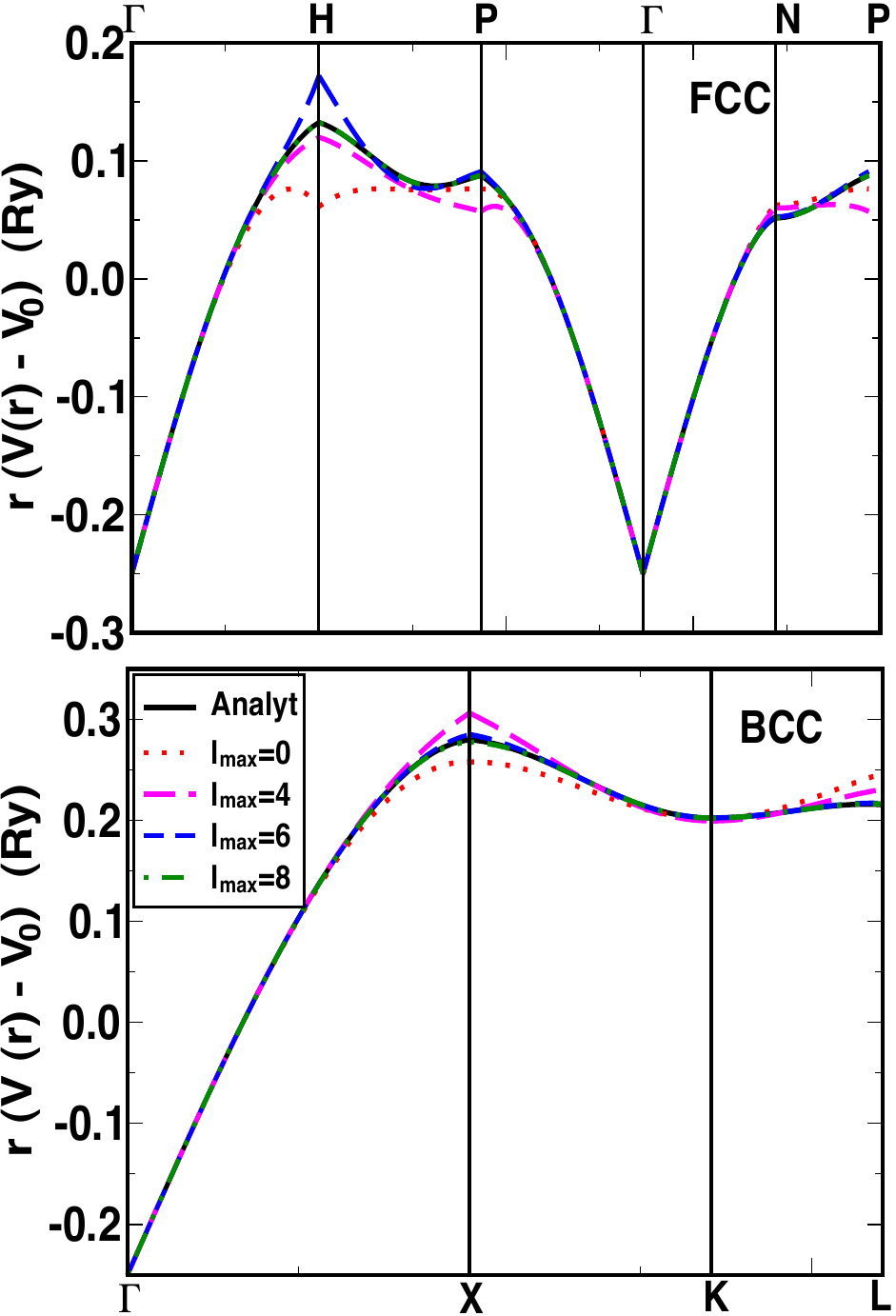}
\caption {(Color online) rV({\bf r}) for various $l_{\text{max}}$ along high-symmetry directions in WS-cells of fcc (top) and bcc (bottom) for Madelung jellium model.}
\label{Vr_vs_l_jell}
\end{figure}

{\par}In Figure~\ref{Vr_vs_l_jell}, we compare the numerical solution of the spatially-dependent potential from our general Eq.~\eqref{eq5} for $l_{\text{max}}=0,4,6,8,10$ to the numerical evaluation of the exact expression \eqref{eq-jellium-pot} for the jellium case for fcc and bcc lattices.
To assess the agreement, we used $15^3$ Gauss points and $8$ neighbor shells to evaluate Eq.~\eqref{eq5}.

\begin{figure}[b]
\centering
\includegraphics[width=6cm]{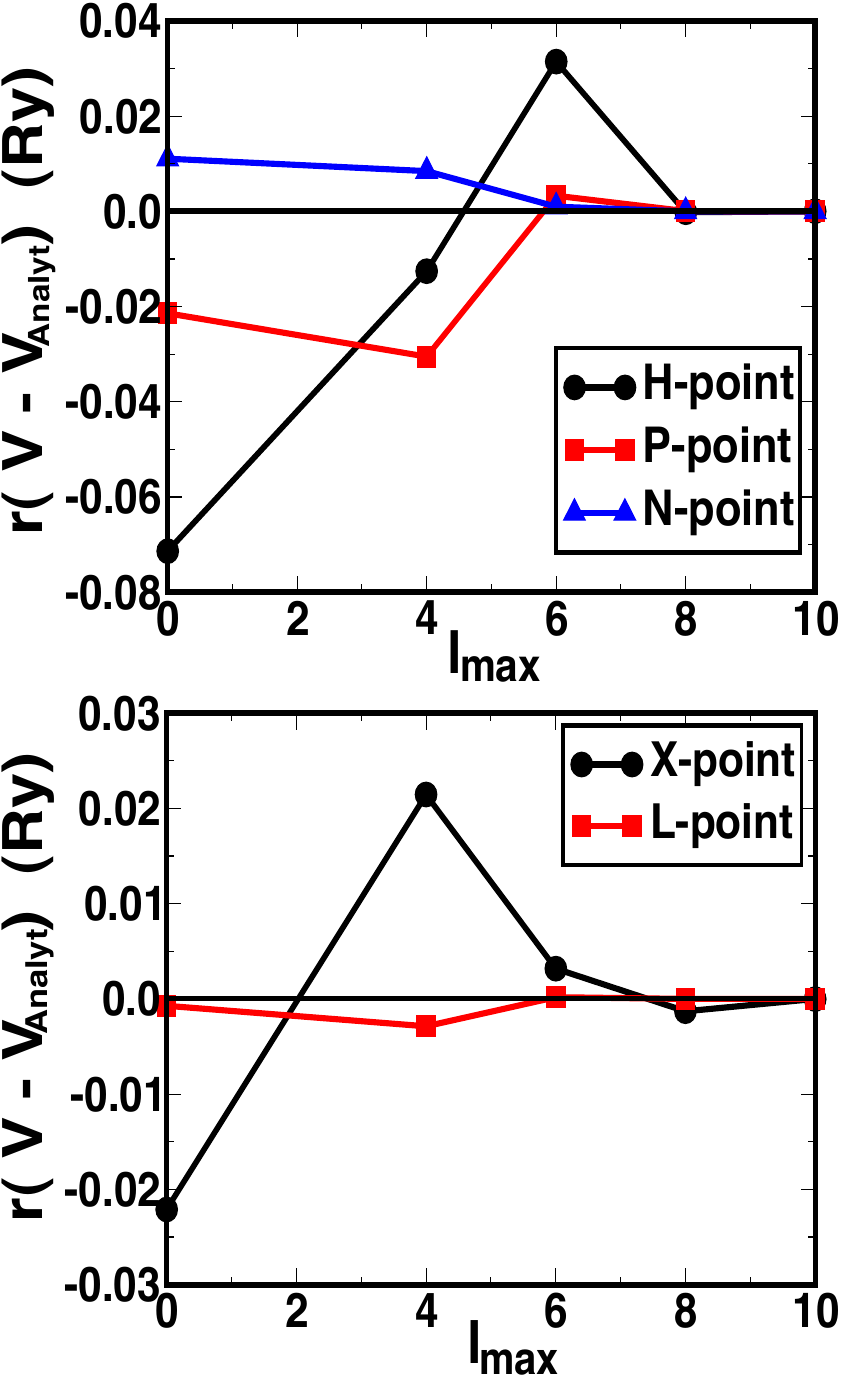}
\caption{(Color online) rV({\bf r}) versus $l_{\text{max}}$ at high-symmetry points in cells of fcc (top) and bcc (bottom) for the Madelung jellium model.}
\label{Vr_vs_lmax_jell}
\end{figure}

{\par}Similar to the van W. Morgan case, the accuracy of the potential for this jellium model varies along the high-symmetry directions, being worse at the H, P point for fcc, and X, L point for bcc case, hence, requiring a higher $L$-sum to approach the analytical closed-form solution, Eq.~\eqref{eq-jellium-pot}. 
Convergence of the potential versus $l_{\text{max}}$ at these points are shown in Fig. \ref{Vr_vs_lmax_jell}, where the NFC are large, see below.
 Unlike previous approaches,\cite{Zhang94,Vitos95,Nicholson2002,Weinberger_book} the present method achieves a much better accuracy even at a lower $l_{\text{max}}$. 
 In contrast to Zhang's\cite{Zhang94} method, which happen to produce fortuitously better potential for $l_{\text{max}}=4$ than $l_{\text{max}}=6$ near the corner of the cell (H-point), the overall quality of our potential improves consistently as $l_{\text{max}}$ is increased. 
 Additionally, in all these other methods, one needs to converge carefully the internal $L^{\text{int}}$-sums; in most cases must be taken up to $l^{\text{int}}_{\text{max}} > 3l^{\text{ext}}_{\text{max}}$,  
 and hence computationally expensive.
However, Hammerling et al.\cite{Hammerling2006} have shown that a multipole approach requires $l^{\text{int}}_{\text{max}} \ge 6l^{\text{ext}}_{\text{max}}$ for the van W. Morgan and Madelung models to achieve accuracy closer to our results.

\begin{figure}[]
\centering
\includegraphics[width=7.5cm]{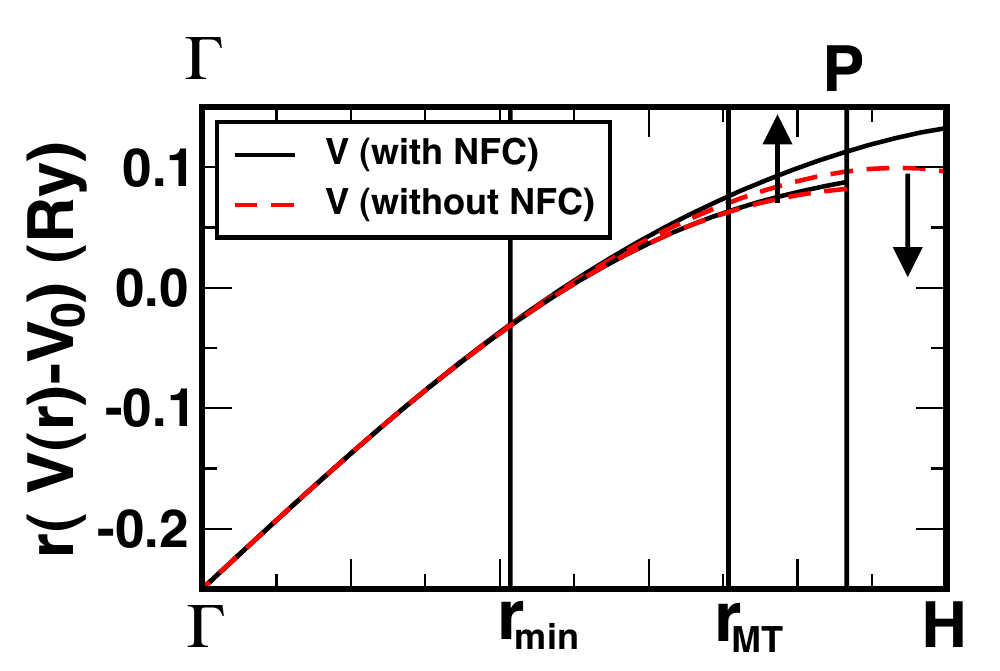}
\caption {(Color online) For fcc, the potential with (without) NFC along $\Gamma-$H and $\Gamma-$P for the Madelung jellium model. Other details are the same as in Fig.~\ref{NFC}.}
\label{NFC-jellium}
\end{figure}

{\par} Again, the NFCs are the reason for a slower rate of convergence near the cell boundary, see Fig.~\ref{NFC-jellium}, where the NFC contribution to the potential for an fcc lattice are shown along the two symmetry directions with $l_{\text{max}}=10$. 
As before, this correction grows only beyond $r_{\text{min}}$ and get significant after $r_{MT}$ as the two densities in Eq.~\eqref{eq5} are identical except outside the central cell where only $\rho^{ex}_L(r) \ne 0$. Unlike the van W. Morgan case, the NFC along both the directions (especially along $\Gamma$-P) in the present case is relatively smaller, reflecting the distinct nature of the two models we have considered.

{\par} Finally, we address the convergence properties of the Coulomb energy for the Madelung problem.
By removing the self-energy arising in the blind application of Eq.~\eqref{eq7} for the Madelung problem, a closed-form solution for the Coulomb energy U (for $N$ unit cells) associated with the potential in Eq.~\eqref{eq-jellium-pot} can be derived, i.e.,
\begin{widetext}
\begin{eqnarray}   \label{eq-jellium-U}
U  = -\left(\frac{NZ^{2}}{r_{asa}}\right) \left(\frac{r_{asa}}{a}\right) 
    \left(
      \frac{4\pi}{\Omega_0\epsilon^2} + \frac{\epsilon}{\sqrt\pi} 
    - \sum_{{\bf R}_n\ne0} \frac{ \emph{erfc}(\epsilon|{\bf R}_n|)}{ |{{\bf R}_n|}}  
    - \frac{4\pi}{\Omega_0} \sum_{ {{\bf K}_{m} \ne 0}} 
      \frac{\emph{exp}({-|{\bf K}_{m}|^2}/4\epsilon^2)}{|{\bf K}_{m}|^2}    
    \right)  .
\end{eqnarray}
\end{widetext}
For convenience, $r_{asa}=(3\Omega_0/4\pi)^{1/3}$ is included, i.e., the radius for a sphere with equivalent unit cell volume $\Omega_0$, i.e., used in the atomic-sphere approximation (ASA).
With this definition, $U/(NZ^{2}r_{asa}^{-1})$ gives exactly $1.8$ for the ASA Madelung problem, whereas the numerical evaluation of Eq.~\eqref{eq-jellium-U} gives $1.79174723$ ($1.79185851$) for fcc (bcc), as found historically.\cite{Skriver1985}
Using the potential and charge density within our Eqs.~\eqref{eq2}-\eqref{eq5}, we can evaluate the integrals for each VP and compare to the results of Eq.~\eqref{eq-jellium-U}.

\begin{figure}[]
\centering
\includegraphics[width=7.5cm]{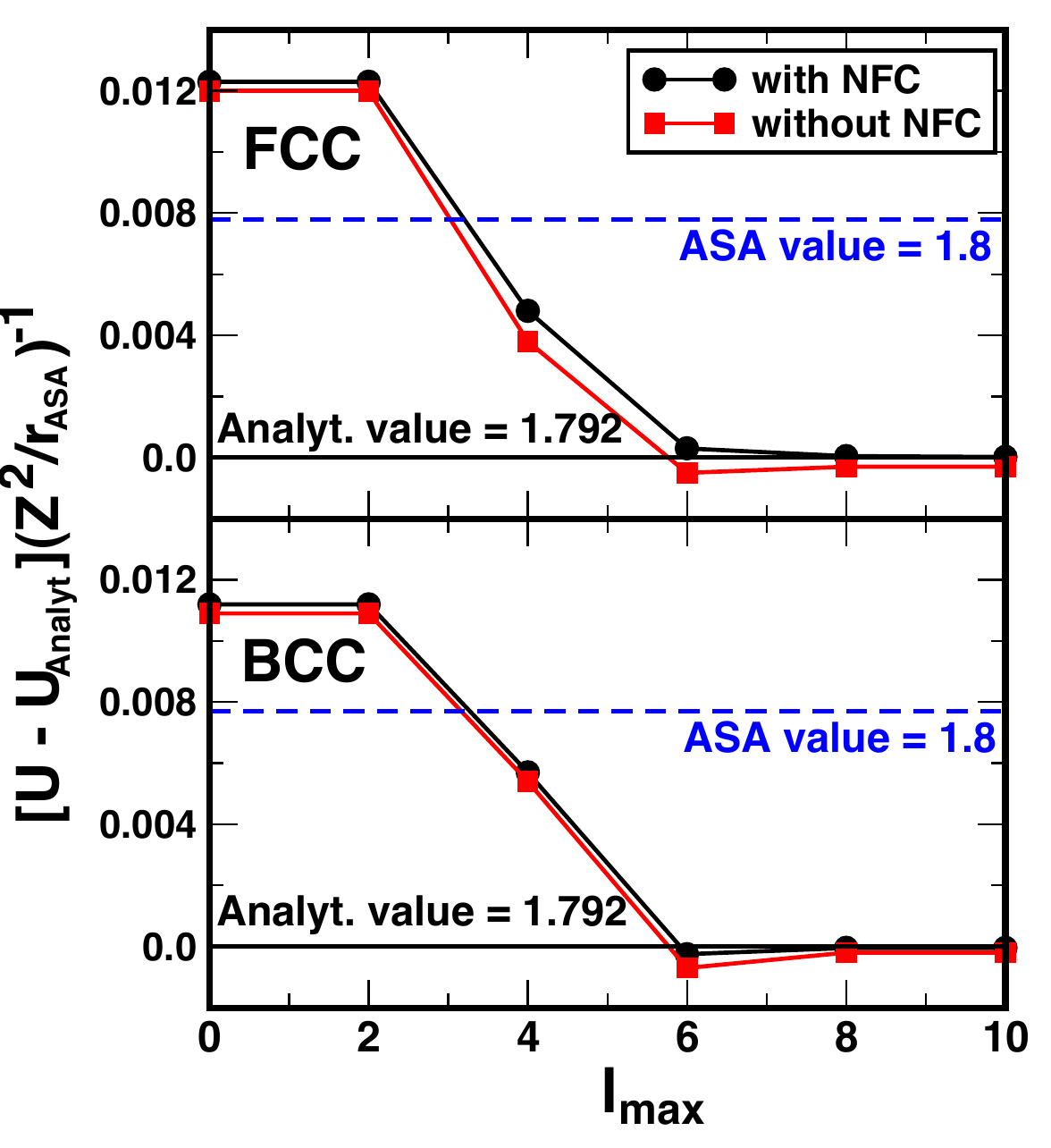}
\caption {(Color online) Coulomb energy for the Madelung problem for fcc and bcc, relative to the results from  Eq.~\eqref{eq-jellium-U}.}
\label{fig-U-madelung}
\end{figure}

{\par}Figure~\ref{fig-U-madelung} shows the convergence of $U$ versus $l_{\text{max}}$ for fcc and bcc lattices, compared to the exact result. 
For the Coulomb energy, the NFC do not have dramatic effects, but there is error without them. 
No systematic error cancellation is possible, e.g., for U$_{\text{fcc}}$-U$_{\text{bcc}}$, which is the well-known Ewald or "muffin-tin" corrections to the ASA structural energies.
The Coulomb energy is almost correct by $l_{\text{max}}=6$ (error at $10^{-6}$ by $l_{\text{max}}=8$), and the convergence is monotonic, unlike when using multipole-based approaches with nested $L$ sums, as shown by Hammerling et al.,\cite{Hammerling2006} where $l^{\text{int}}_{\text{max}} \ge 6l^{\text{ext}}_{\text{max}}$ to achieve $10^{-6}$ accuracy comparable to our results without internal sums, which are very slowly convergent and numerically costly.

\subsection{General Comments}
\label{sec_results_general}

{\par} Our isoparametric integration avoids conditionally convergent summations, required in previous approaches, and provides a significantly more accurate and faster method for solving Poisson's equation, as detailed by the two cases.
For molecular systems, a finite sum over atoms is required.
For extended, solid-state systems, it also avoids FFTs, a limiting factor for large-atom cell calculations.
In general, the present method is at least $10(l^{\text{int}}+1)^2  \text{N}_{\text{VP}}$ times faster than any of the existing schemes.\cite{Gonis91,Schadler92,Nicholson2002} 
The factor $(l^{\text{int}}+1)^2$ comes from an additional internal $L$-sum (typically $l^{\text{int}} \sim 6 l^{\text{ext}}$), and the factor $10$ is from use of isoparametric integration versus shape functions, if used. 
In particular, for a system with $\text{N}_{\text{VP}}$ sublattices, $l^{\text{ext}}\sim 8-10$ will provide $\sim$$10^4 \text{N}_{\text{VP}}$ speed up.
A direct comparison of CPU timings was detailed recently\cite{Alam2011} and shows that isoparametric integration is $10^5$ faster and $10^7$ more accurate than that using shape functions.

\section{Summary}
\label{sec-summary}
{\par} We have resolved the longstanding problem of an accurate, fast and efficient numerical solution of Poisson equation for electronic-structure codes with site-centered basis-sets. 
In particular, a proper calculation of the intercell potential has been developed that avoids troublesome multipole-type techniques that are conditionally convergent and we include accurately the correction term from the near cells, the so-called Near-Field Correction, where we have developed a physically intuitive and fast method to evaluate this correction also without multipoles.
The method provides machine-precision for potentials and Coulomb energy for systems described by arbitrarily-shaped, convex, space-filling VP, eliminates previous computational bottlenecks and convergence issues by employing isoparametric integration, scales as O(N$_{\text{VP}}$) and is easily parallelized. 
The method also avoids FFTs that do not scale well to very large cells.
The method works for periodic solids, molecules (using extended VP) and materials containing imperfections or disorder. 
The general applicability and accuracy of the method was proved via two rigorous, analytic models that traverse from localized to extended densities.

\section{Acknowledgements} 
Research sponsored by the U.S. Department of Energy, Office of Basic Energy Science, Division of Materials Science and Engineering Division from contracts with DDJ (DEFG02-03ER46026) and seed funding with Ames Laboratory, which is operated for DOE by Iowa State University under contract DE-AC02-07CH11358; 
from the ``Center for Defect Physics'', an Energy Frontier Research Center, for DDJ to support a student who helped develop numerical integration method (Ref.~\onlinecite{Alam2011}) used here and in our EFRC's code. Work performed by  BGW was under the auspices of the U.S. DOE by Lawrence Livermore National Laboratory under Contract DE-AC52-07NA27344. We also benefited from discussion with W.A.~Shelton in our DOE/BES Computational Materials and Chemical Sciences Network, and D.M.C.~Nicholson in the EFRC, to reproduce their method and results in Ref.~\onlinecite{Nicholson2002}.

\vspace{-0.5cm}

\end{document}